\documentclass{emulateapj}

\shorttitle{The UV Galaxy Luminosity Function at low redshift}
\shortauthors{Treyer et al.}

\begin{document}

\title{The Ultraviolet Galaxy Luminosity Function from GALEX data: 
Color Dependent Evolution at Low Redshift}

\author{Marie Treyer\altaffilmark{1,2}, Ted K. Wyder\altaffilmark{1}, 
David Schiminovich\altaffilmark{1},
St\'ephane Arnouts\altaffilmark{2},
Tam\'as Budav\'ari\altaffilmark{4}, 
Bruno Milliard\altaffilmark{2},
Tom A. Barlow\altaffilmark{1},
Luciana Bianchi\altaffilmark{4},
Yong-Ik Byun\altaffilmark{3}, 
Jos\'e Donas\altaffilmark{2},
Karl Forster\altaffilmark{1},
Peter G. Friedman\altaffilmark{1},
Timothy M. Heckman\altaffilmark{4},
Patrick N. Jelinsky\altaffilmark{5},
Young-Wook Lee\altaffilmark{3},
Barry F. Madore\altaffilmark{6},
Roger F. Malina\altaffilmark{2},
D. Christopher Martin\altaffilmark{1},
Patrick Morrissey\altaffilmark{1},
Susan G. Neff\altaffilmark{7},
R. Michael Rich\altaffilmark{8},
Oswald H. W. Siegmund\altaffilmark{5},
Todd Small\altaffilmark{1},
Alex S. Szalay\altaffilmark{4}, and
Barry Y. Welsh\altaffilmark{5}}

\altaffiltext{1}{California Institute of Technology, MC 405-47, 1200 East
California Boulevard, Pasadena, CA 91125; treyer@srl.caltech.edu}
\altaffiltext{2}{Laboratoire d'Astrophysique de Marseille, BP 8, Traverse
du Siphon, 13376 Marseille Cedex 12, France}
\altaffiltext{3}{Center for Space Astrophysics, Yonsei University, Seoul
120-749, Korea}
\altaffiltext{4}{Department of Physics and Astronomy, The Johns Hopkins
University, Homewood Campus, Baltimore, MD 21218}
\altaffiltext{5}{Space Sciences Laboratory, University of California at
Berkeley, 601 Campbell Hall, Berkeley, CA 94720}
\altaffiltext{6}{Observatories of the Carnegie Institution of Washington,
813 Santa Barbara St., Pasadena, CA 91101}
\altaffiltext{7}{Laboratory for Astronomy and Solar Physics, NASA Goddard
Space Flight Center, Greenbelt, MD 20771}
\altaffiltext{8}{Department of Physics and Astronomy, University of
California, Los Angeles, CA 90095}

\begin{abstract}
We present measurements of the FUV (1530\AA) and NUV (2310\AA) galaxy luminosity functions 
(LF) at low redshift ($z\le0.2$) from GALEX observations matched to the 2dF 
Galaxy Redshift Survey.
We split our FUV and NUV samples into two UV$-b_j$ color bins and two redshift bins. 
As observed at optical wavelengths, the local LF of the bluest galaxies tend to have 
steeper faint end slopes and fainter characteristic magnitudes M$_{\ast}$ 
than the reddest subsamples. We find evidence for color dependent evolution 
at very low redshift in both bands, with bright blue galaxies becoming dominant
in the highest redshift bin. The evolution of the total LF is consistent with 
an $\sim 0.3$ magnitude brightening between $z\sim$ 0 and 0.13, in agreement 
with the first analysis of deeper GALEX fields probing adjacent and higher redshifts. 
\end{abstract}

\section{Introduction}

The importance of estimating luminosity functions (LF) as a function
of redshift, color, environment, wavelength, etc., to understand galaxy 
formation and evolution, has been well emphasized in the literature.  
Multivariate luminosity functions are crucial to constrain theoretical
models on small scales where the physics is most complex. The shape
of the galaxy LFs results from non trivial physical processes
(Benson et al.~2003, Binney 2004) and understanding
their evolution is a challenging task for numerical simulations 
(e.g. Nagamine et al.~2001).
The Galaxy Evolution Explorer (GALEX) mission has recently 
opened a new field of constraints by
allowing such statistical measures as LFs to be performed in the 
rest-frame Ultraviolet (UV) in what is now considered the low redshift 
Universe ($z<1-2$), where most of the evolution of galaxies is thought 
to have taken place. 

In this paper, we use data from the GALEX All-Sky Imaging Survey 
matched to spectroscopic data from the 2dF Galaxy Redshift Survey to investigate
the color properties of local UV selected galaxies, and the color dependence
and evolution of their LF at $very$ low redshift ($z<0.2$). Whereas much effort
has been devoted to estimate the evolution of galaxy light over the widest 
and furthest range of redshift possible, very few surveys have allowed to probe 
evolution at recent epochs (Loveday 2004), and none so far at UV wavelengths.   
This work is a companion paper to Wyder et al.~(2004) who present the 
total LFs of FUV and NUV selected galaxies in the local ($z\le 0.1$)
Universe (hereafter Paper I). It also complements the work of 
Arnouts et al. (2004), Schiminovich et al. (2004) and Budav\'ari et al. (2004) 
who quantify the evolution of the UV LF at higher redshift using the GALEX 
Medium and Deep Imaging Surveys. 

The data are summarized in Section 2. 
Section 3 presents the color properties of the samples. We estimate and
discuss the dependence of the UV LFs on galaxy color and redshift at recent 
epoch in Section 4. We assumed a flat $\Lambda$CDM cosmology with 
H$_0=70~{\rm km~ s^{-1} Mpc^{-1}}$, $\Omega_M=0.3$ and
$\Omega_{\Lambda}=0.7$.

\section{Data}

The GALEX field-of-view has a diameter of 1.2\arcdeg and each
pointing is imaged simultaneously in both the FUV and NUV bands
with effective wavelengths of ${\rm 1530~\AA}$ and ${\rm 2310~\AA}$,
respectively. 
The GALEX instruments and mission are described by \citet{morissey04}
and \citet{martin04}.

The specific data analyzed in this paper are presented in Paper I.
In brief, they consist of 133 GALEX All-Sky Imaging Survey
pointings overlapping the 2dF Galaxy Redshift Survey \citep{colless01}
in the South Galactic Pole region. 
As the NUV images are substantially deeper than the FUV, we
used the NUV images for detection and measured the FUV flux in the
same aperture as for the NUV.
The apparent magnitudes were corrected for foreground extinction
using the \citet{schlegel98} reddening maps and assuming the
extinction law of \citet{cardelli89}. 

The GALEX catalogs were matched with the 2dF catalog using
a search radius of $6\arcsec$.
We restricted the sample to areas with effective exposure 
times $t_{eff}>60$ sec, and removed sources with
magnitude errors greater than 0.4 or contaminated by
artifacts from bright stars. 
We also removed overlap regions by restricting the coverage to 
the inner $0.45\arcdeg$ of each field.
Finally, we excluded GALEX sources where the
2dF redshift completeness was less than 80\% as well as areas of the
sky excluded from the 2dF input catalog.  
Our final working area of GALEX-2dF overlap is 56.73 deg$^2$ in both bands.

We applied a faint magnitude limit of 20 in both bands corresponding to the 
bluest $UV-b_j$ color observed and beyond which the redshift completeness begins 
to drop. (N.B.: $b_j$ was shifted by $-0.051$ into the AB system. All quoted 
magnitudes are AB).
We also applied a bright magnitude limit of 17 to avoid extended
sources with potential photometric problems. We visually inspected the resulting 
catalog of 2dF spectra and further removed 27 objects with broad emission lines 
(the obvious AGN/QSO's). Our final catalogs consist of 1292 FUV-selected galaxies 
and 1869 NUV-selected galaxies with $17\le UV \le 20$ and available $b_j$ magnitude 
and 2dF redshift (with 2dF redshift quality parameter $\ge 3$).
Both redshift distributions extends to $z \sim 0.25$ (Paper I).
FUV fluxes are available for all but one of the galaxies in 
the NUV-selected sample, although uncertainties can be quite 
large for the faintest objects.

\section{Color properties}

\begin{figure}
\plotone{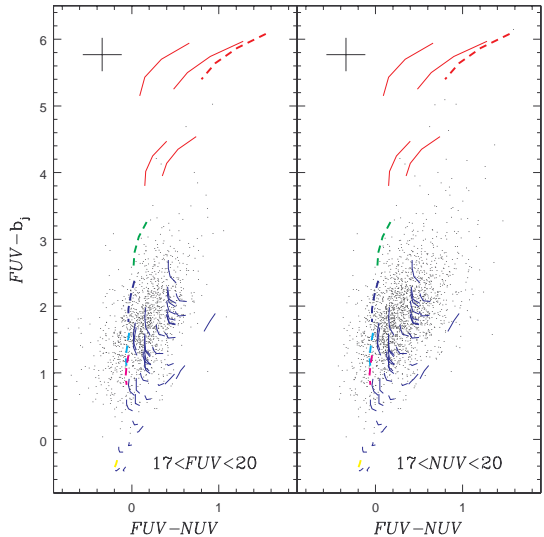}
\caption{Color-color diagrams: $FUV-NUV$ vs $FUV-b_j$ for the FUV (left panel)
and the NUV (right panel) selected samples respectively. The thin solid lines
are a set of models by Bruzual \& Charlot (2003) covering a range of star-formation
histories, ages, metallicities and extinctions. The thick dashed lines are dust free
models by Poggianti (1997), from Elliptical to Starburst. 
\label{colcol}}
\end{figure}

Figure \ref{colcol} shows color-color diagrams ($FUV-NUV$ vs $FUV-b_j$)
for the FUV and NUV selected samples (left and right panel, respectively). 
Typical error bars are shown in the upper left corners. 
The thin lines are a set of 42 spectral energy distribution (SED) models 
by Bruzual \& Charlot (2003) (hereafter BC), from old ellipticals 
(in red) to young irregular galaxies, assuming a range of star-formation 
histories ($\tau=1$ Gyr to $\infty$), ages (0.1 to 12 Gyr), metallicities 
(0.2 $Z_{\odot}$ to $Z_{\odot}$) and extinctions ($A_V=0$ to 1.5). 
The thick dashed lines are dust free SED models by Poggianti (1997). 
The tracks show the color evolution of the various galaxy types from 
$z=0$ to 0.25 (roughly counter-clockwise). 

The bulk of both samples is consistent with bona fide late-type galaxies,
with only a handful of objects displaying elliptical-like colors.
We do not observe any source with ``$UV$-optical excess'' colors (bluer than the 
bluest model) such as were observed in the FOCA sample (Treyer et al.~1998, 
Sullivan et al.~2000), suggesting that the so-called excess was to be found in the 
estimate of the UV flux of these objects rather than in some physical explanation. 
On the other hand we find a small fraction of galaxies with $FUV-NUV$ colors somewhat 
bluer than any model ($\sim 15\%$ in the FUV selected sample, $\sim 7 \%$ in the NUV 
selected sample). This is partly due to newly measured, small but opposite offsets 
in the FUV and NUV calibrations, adding up to +0.13 in $FUV-NUV$ (new in-flight 
calibrations, P. Morissey, private communication), not applied in the current
catalogs. 
But shortcomings in the models at UV wavelengths cannot be ruled out. 
Despite the wide coverage in parameter space, the data are not fully accounted for 
(e.g. the reddest sources in both $FUV-NUV$ and $FUV-b_j$). Fig.~\ref{colcol} suggests 
that UV selected galaxies may require alternative model ingredients (e.g. more erratic 
star-formation histories as suggested by Sullivan et al.~2003).

\begin{figure}
\plotone{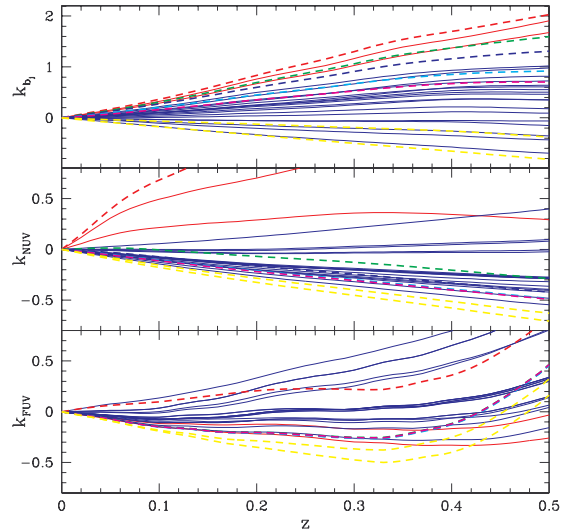}
\caption{Model $k$-corrections in the $b_j$, FUV and NUV bands. The thin solid 
lines show our set of BC models and the thick dashed lines the Poggianti 
SEDs, as in Fig.~\ref{colcol}. 
\label{kcorr}}
\end{figure}

We assign a best fit SED (drawn from the combined and interpolated BC and 
Poggianti libraries) to each galaxy using its redshift and two independent colors 
($FUV-NUV$ and $FUV-b_j$).
Figure \ref{kcorr} shows the model $k$-corrections
in the $b_j$, FUV and NUV bands (thin solid lines for BC and thick dashed lines for Poggianti). 
UV $k$-corrections are small enough for the late-type, low-redshift 
galaxies dominating the sample (as opposed to $k$-corrections in optical bands)
that uncertainties on the assigned SEDs have negligible impact on 
the resulting LFs.

We can also fit a power-law through the 2 bandpasses and derive a spectral index 
for each galaxy ($\beta=(FUV-NUV)/0.425 -2$, assuming $F_{\lambda}\propto \lambda^{\beta}$). 
The mean spectral indices of the FUV and NUV selected samples 
are $\langle \beta\rangle \sim -1.6$ and $-1.3$ respectively. 
The UV slope $\beta$ has been shown to be a good tracer of dust
attenuation in galaxies undergoing a strong starburst, but not so reliably in
others (Bell 2002, Kong et al.~2004). In particular, Buat et al.~(2004) found that 
local NUV selected galaxies tend to have lower attenuations than starburts 
for a given slope. 
As our two samples exhibit slopes corresponding to low attenuations in pure starburst 
galaxies, we conclude that they must be very little affected by dust. (See Paper I for 
a quantitative discussion). 

\section{Color and $z$ dependent luminosity functions}

We split both samples into two rest-frame color bins: 
$(FUV-b_j)_0 \le$ and $\ge 2$ (roughly corresponding to an Sb spectral type and
to the mean of the color distribution), referred to as blue and red subsamples 
respectively. We further split each of these subsamples into two redshift bins: 
$z\le 0.1$ and $0.1\le z\le 0.2$, referred to as low and `higher' redshift 
subsamples respectively. The mean redshifts in these two bins are 0.055 and 0.12 
respectively for the FUV sample, and 0.057 and 0.13 for the NUV sample. 

As in Paper I, the redshift completeness is defined as the ratio of UV objects 
matched to a 2dFGRS counterpart with (high quality) redshift
to the total number of UV galaxies in a given magnitude bin,
as computed by Xu et al.~(2004) using the SDSS star/galaxy 
separation criteria in overlap regions. 
It is roughly constant over the range $17 \le UV \le 20$ in both bands, with mean 
values of $\sim$ 79\% in the NUV and 92\% in the FUV (cf Fig.~1 in Paper I).
We computed binned luminosity functions using the traditional V$_{max}$ estimator
\citep{felten76} in the 12 subsamples (blue, red, and total at low and higher $z$
in both bands),
and derived best fit Schechter functions \citep{schechter76} for each of them.
Since the magnitude range in the highest redshift slice is too bright and narrow to 
constrain the slope meaningfully, we fixed it to the low redshift value in each case. 

\begin{deluxetable}{cccccc}
\tablecolumns{6}
\tablewidth{0pc}
\tablecaption{Schechter function parameters ($\phi_{\ast}$ is in Mpc$^{-3}$).\label{params}}
\tablehead{
\colhead{Band} & \colhead{$z$} & \colhead{Type} & \colhead{M$_{\ast}$} & \colhead{$-\alpha$} & 
\colhead{$\log{\phi_{\ast}}$} 
}
\startdata
FUV&0.0-0.1& All  &$-18.04 \pm   .11 $&$  1.22 \pm   .07 $&$ -2.37 \pm   .06  $\\ 
  -&0.1-0.2&   -  &$-18.31 \pm   .06 $&-                  &$ -2.33 \pm   .07  $\\ 
  -&0.0-0.1& Blue &$-17.89 \pm   .15 $&$  1.29 \pm   .09 $&$ -2.54 \pm   .09  $\\ 
  -&0.1-0.2&   -  &$-18.44 \pm   .09 $&-                  &$ -2.73 \pm   .10  $\\ 
  -&0.0-0.1& Red  &$-18.19 \pm   .22 $&$  1.10 \pm   .17 $&$ -2.80 \pm   .12  $\\ 
  -&0.1-0.2&   -  &$-18.15 \pm   .08 $&-                  &$ -2.50 \pm   .10  $\\ 
NUV&0.0-0.1& All  &$-18.23 \pm   .11 $&$  1.16 \pm   .07 $&$ -2.26 \pm   .06  $\\ 
  -&0.1-0.2&   -  &$-18.53 \pm   .05 $&-                  &$ -2.27 \pm   .05  $\\ 
  -&0.0-0.1& Blue &$-18.04 \pm   .16 $&$  1.25 \pm   .10 $&$ -2.48 \pm   .09  $\\ 
  -&0.1-0.2&   -  &$-18.71 \pm   .08 $&-                  &$ -2.82 \pm   .08  $\\ 
  -&0.0-0.1& Red  &$-18.28 \pm   .18 $&$  0.97 \pm   .14 $&$ -2.56 \pm   .09  $\\ 
  -&0.1-0.2&   -  &$-18.36 \pm   .06 $&-                  &$ -2.37 \pm   .06  $\\ 
\enddata
\end{deluxetable}

\begin{figure*}
\plotone{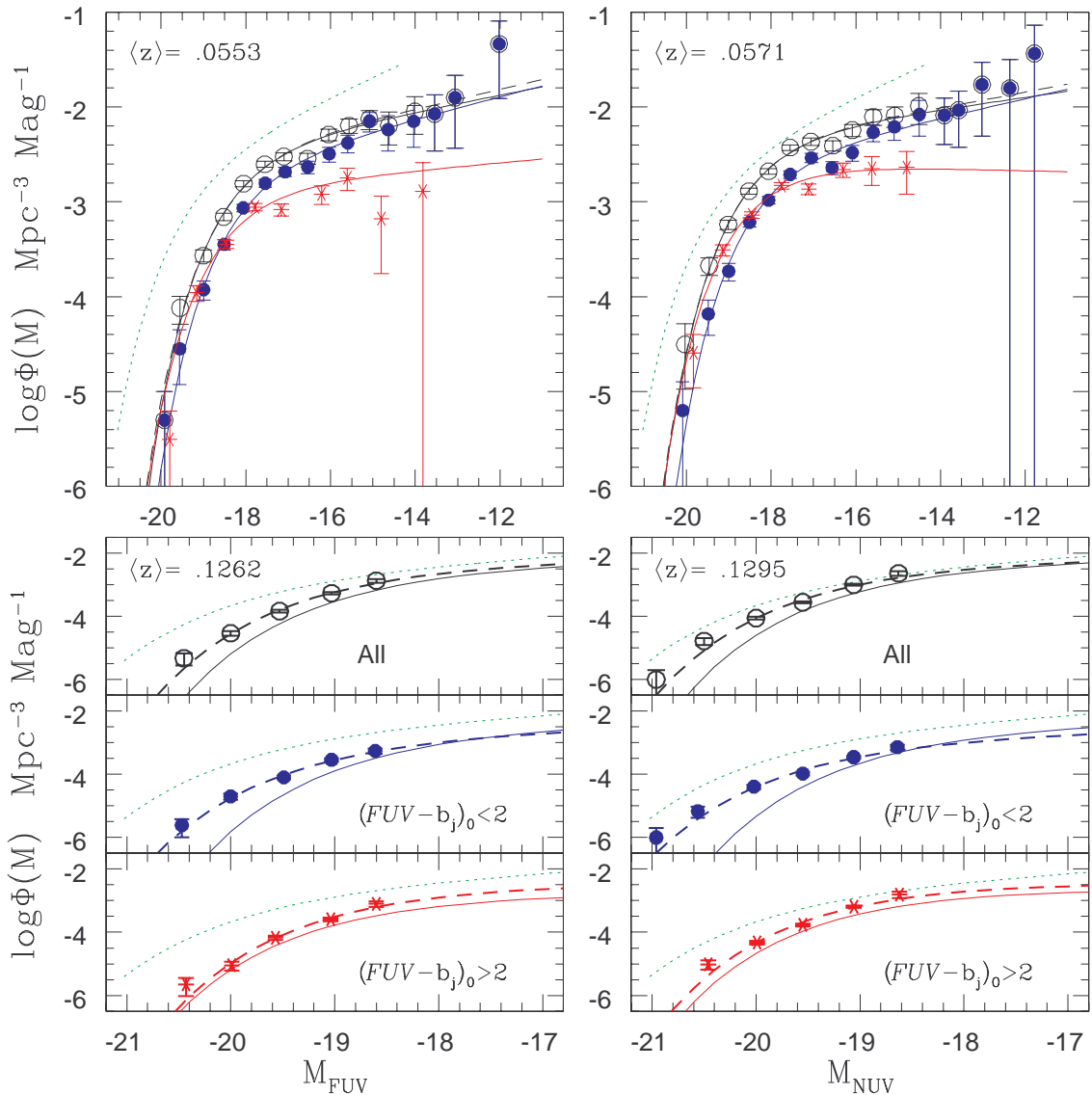}
\caption{{\bf Top panels :} The FUV and NUV LFs at $z\le 0.1$ (left and right
panel, respectively). The blue and red
lines (and associated filled circles and asteriks) represent the blue and red 
subsamples respectively. The black solid lines (and associated empty circles)
show the LFs of the full samples (Paper I), while the dotted lines are the sum
of the blue and red Schechter functions. The green dotted line is the 
FOCA 2000\AA\ LF at $\bar z \sim 0.15$ (Sullivan et al.~2000). {\bf Bottom panels :}
The FUV and NUV LFs at $0.1 \le z \le 0.2$. Same symbols and lines as above.
\label{lfs}}
\end{figure*}

Results are shown in Fig.~\ref{lfs} 
and the best-fit Schechter function parameters are listed in Table \ref{params}.  
The FOCA 2000\AA\ LF (Sullivan et al.~2000) at a mean redshift ${\bar z}\sim0.15$ is also 
plotted for comparison (converted to the AB magnitude system and to our H$_0$ 
value but without accounting for the difference in bandpasses). 
We refer to Paper I for a discussion on the discrepancy between the FOCA and GALEX LFs.
As observed at optical wavelengths (Madgwick et al.~2002, Nakamura et al.~2003),
the local LFs of the bluest galaxies tend to have steeper faint end slopes and 
fainter characteristic magnitudes M$_{\ast}$ than the reddest subsamples (although
our definition of red would qualify as blue in optically selected samples).
The fact that the latest types are observed to fainter absolute magnitudes 
than the earlier types induces a small bias in the estimate of the faint end
slope of the full sample. The dotted lines in the upper panels of Fig.~\ref{lfs} 
show that the sum of the blue and red LFs is slightly steeper at the faint end than estimated 
for the galaxy population as a whole, due to the shortage of red galaxies in the faintest bins.

The total FUV and NUV LFs in the highest redshift bin (lower panels in Fig.~\ref{lfs}) 
are both consistent with $\sim 0.3$ magnitude brightening in M$_{\ast}$ with respect to 
the LFs at $z<0.1$. This evolution is quite similar to that found in optically selected
samples at the same redshifts (Loveday 2004).
Evolution can also be inferred from the steeply rising V/V$_{max}$ 
distribution of both $z$-unlimited samples (with $\langle$V/V$_{max}\rangle =0.54$ in both 
bands instead of the 0.5 expected from a non evolving population). As the redshift
distributions show two strong peaks around 0.06 and 0.11 (Paper I), the appearance
of evolution could be attributed to large scale structure effects, but removing these 
two structures from the samples did not affect our results.

Evolution also seems to occur predominantly in the blue subsamples, with
blue galaxies dominating the bright end of the LF in the highest $z$ bin.
We checked that no color trend was observed as a function of $b_j$, as might
be expected if the 2dFGRS was strongly biased against, e.g., faint blue low surface 
brightness galaxies (LSBG), making bright blue galaxies seem dominant at
higher redshift. The issue of missing galaxies in the 2dFGRS,
in particular LSBGs, has been addressed in a number of papers, 
most recently by Cross et al.~(2004). The authors find that the survey 
does indeed miss $\sim 15\%$ of the galaxy population, of which $6\%$
only are identified as LSBGs. If this particular class of objects is unlikely 
to affect our results, the remaining missing galaxies may bias our analysis, 
should their redshift distribution look different from the one actually measured. 
A truly UV-selected spectroscopic survey is underway to address this question. 

The present results are consistent with the FUV analysis of a much deeper GALEX field 
probing adjacent and higher redshifts (Arnouts et al.~2004, Schiminovich et al.~2004).
They are also in rough agreement with the rate of evolution derived 
from a photometric-redshift study of GALEX/SDSS data at low $z$
(Budav\'ari et al.~2004), although our FUV LF in the range $0.1 \le z \le 0.2$ 
seems brighter than the photo-$z$ estimate at overlapping $z$. Both 
analysis agree very well in the NUV however (we postpone attempts at 
explanations for the FUV discrepancy until more comparable datasets are available). 
We note that evolution seems to be limited by the observed 
galaxy number counts, especially in the FUV band (Xu et al.~2004), so that the rate 
of evolution detected  at $z\le 1$ in the GALEX data is expected to slow down. 
This appears to be the case based on a photometric-redshift analysis of the HDF North 
and South probing $1.75\le z\le 3.4$ (Arnouts et al.~2004). 

\acknowledgments   
MT thanks the GALEX team at Caltech for its great hospitality.
GALEX is a NASA Small Explorer, launched in April 2003. We gratefully acknowledge 
NASA's support for its construction, operation, and science analysis, as well as the
cooperation of the French Centre National d'Etudes Spatiales and of the Korean Ministry 
of Science and Technology.


\begin{thebibliography}

\bibitem[Arnouts et al.(2004)]{arnouts04} Arnouts S., et al.~2004, \apjl, submitted
\bibitem[Bell(2002)]{2002ApJ...577..150B} Bell , E.~F.~2002, \apj, 577, 150
\bibitem[Benson et al.(2003)]{2003ApJ...599...38B} Benson, A.~J., Bower, R.~G., Frenk, C.~S., Lacey, C.~G., Baugh, C.~M., \& Cole, S.\ 2003, \apj, 599, 38 
\bibitem[Binney(2004)]{2004MNRAS.347.1093B} Binney, J.\ 2004, \mnras, 347, 1093 
\bibitem[Bruzual \& Charlot(2003)]{BC03} Bruzual, G. \& Charlot, S. 2003, \mnras, 344, 1000 
\bibitem[Buat et al.(2004)]{buat04} Buat V. et al.~2004, \apjl, submitted
\bibitem[Budav\'ari et al.(2004)]{budavari04} Budav\'ari T., et al.~2004, \apjl, submitted
\bibitem[Cardelli et al.(1989)]{cardelli89} Cardelli, J. A., Clayton, G. C.,
\& Mathis, J. S. 1989, \apj, 345, 245
\bibitem[Colless et al.(2001)]{colless01} Colless, M., et al.~2001, \mnras, 328, 1039
\bibitem[Cross et al. (2004)]{cross04} Cross, N.J.G., et al.~2004, \mnras, 349, 576 
\bibitem[Felten(1976)]{felten76} Felten, J. E. 1976, \apj, 207, 700
\bibitem[Kong et al.(2004)]{kong04} Kong, X., Charlot, S., Brinchmann, J., \& Fall, M. 2004,
astro-ph/0312474
\bibitem[Loveday(2004)]{loveday04} Loveday, J. 2004, \mnras, 347, 601
\bibitem[Madgwick et al.(2002)]{2002MNRAS.333..133M} Madgwick, D.~S., et 
al.\ 2002, \mnras, 333, 133 
\bibitem[Martin et al.(2004)]{martin04} Martin, D. C. et al.~2004, \apjl, submitted
\bibitem[Morissey et al.(2004)]{morissey04} Morissey, P. et al.~2004, \apjl, submitted
\bibitem[Nakamura et al.(2003)]{2003AJ....125.1682N} Nakamura, O., et al. \ 2003, \aj, 125, 1682 
\bibitem[Nagamine, Fukugita, Cen, \& Ostriker(2001)]{2001MNRAS.327L..10N} 
Nagamine, K., Fukugita, M., Cen, R., \& Ostriker, J.~P.\ 2001, \mnras, 327, L10 
\bibitem[Poggianti(1997)]{1997A&AS..122..399P} Poggianti, B.~M. 1997, \aaps, 122, 399
\bibitem[Schechter(1976)]{schechter76} Schechter, P. 1976, \apj, 203, 297
\bibitem[Schiminovich et al.(2004)]{schimi04} Schiminovich, D., et al.~2004, \apjl, submitted
\bibitem[Schlegel et al.(1998)]{schlegel98} Schlegel, D. J., Finkbeiner, D. P., \& David, M. 1998, \apj, 500, 525
\bibitem[Sullivan et al.(2000)]{sullivan00} Sullivan, M., Treyer, M. A., Ellis, R. S., Bridges, T., Milliard, B., \& Donas, J. 2000, \mnras, 312, 442
\bibitem[Sullivan et al.(2004)]{sullivan04} Sullivan, M., Treyer, M. A., Ellis, R. S., Mobasher, B. 2004, \mnras, 350, 21
\bibitem[Treyer et al.(1998)]{treyer98} Treyer, M. A., Ellis, R. S., Milliard, B., Donas, J., Bridges, T. J. 1998, \mnras, 300, 303
\bibitem[Wyder, T. et al.(2004)]{wyder04} Wyder, T. K., et al.~2004, \apjl, submitted
\bibitem[Xu et al.(2004)]{xu04} Xu, C.K., et al.~2004, \apjl, submitted

\end{thebibliography}
\end{document}